\documentclass[aps,prl,amsmath,amsfonts,amssymb,nofootinbib,twocolumn,showpacs]{revtex4-1}

\usepackage{graphicx}
\usepackage{eucal}

\begin{document}
\title{Efficiency at maximum power of low dissipation Carnot engines}

\author{Massimiliano Esposito}
\affiliation{Center for Nonlinear Phenomena and Complex Systems,
Universit\'e Libre de Bruxelles, CP 231, Campus Plaine, B-1050 Brussels, Belgium.}
\author{Ryoichi Kawai}
\affiliation{Department of Physics, University of Alabama at Birmingham, 1300 University Blvd. Birmingham, AL 35294-1170, USA.}
\author{Katja Lindenberg}
\affiliation{Department of Chemistry and Biochemistry and BioCircuits Institute, University of California,
San Diego, La Jolla, CA 92093-0340, USA.}
\author{Christian Van den Broeck}
\affiliation{Hasselt University, B-3590 Diepenbeek, Belgium.}

\date{\today}


\begin{abstract}
We study the efficiency at maximum power, $\eta^*$, of engines performing finite-time Carnot cycles between a hot and a cold reservoir at temperatures $T_h$ and $T_c$, respectively. For engines reaching Carnot efficiency $\eta_C=1-T_c/T_h$ in the reversible limit (long cycle time, zero dissipation), we find in the limit of low dissipation that $\eta^*$ is bounded from above by $\eta_C/(2-\eta_C)$ and from below by $\eta_C/2$. These bounds are reached when the ratio of the dissipation during the cold and hot isothermal phases tend respectively to zero or infinity. For symmetric dissipation (ratio one) the Curzon-Ahlborn efficiency $\eta_{CA}=1-\sqrt{T_c/T_h}$ is recovered.
\end{abstract}

\maketitle


Thermal machines performing Carnot cycles transform a certain amount of heat $Q_h$ from a hot reservoir at temperature $T_h$ into an amount of work $-W$, with the remaining energy being evacuated as heat $Q_c=-Q_h-W$ to a cold reservoir at temperature $T_c$. We adopted here the usual convention that heat and work absorbed by the system are positive. By assuming that there is no perpetuum mobile of the second kind, more precisely that heat does not spontaneously flow from a cold to a hot reservoir, Carnot was able to show that the efficiency of the heat-work transformation
\begin{eqnarray}
\eta = -\frac{W}{Q_h} = 1+\frac{Q_c}{Q_h} \label{EffiDef} 
\end{eqnarray}
is universally bounded by a maximum value, the so-called Carnot efficiency 
\begin{eqnarray}
\eta_C=1-\frac{T_c}{T_h} \label{CarnotEff}.
\end{eqnarray}
This insight lies at the heart of thermodynamics, since it led Clausius to the introduction of the  entropy, the state function which is central for the formulation of the Second Law. The entropy change of a system is given by $\Delta S=\int_{qs}\bar{d}Q/T$, where the integral is over the infinitesimal amounts of absorbed heat $\bar{d}Q$ for a quasi-static transformation of the system (i.e., a succession of equilibrium states) \cite{callen}.   
The total entropy production during a Carnot cycle is given by 
\begin{eqnarray}
\Delta S_{tot} = -\frac{Q_h}{T_h}-\frac{Q_c}{T_c} = (\eta_C-\eta) \; \frac{Q_h}{T_c} \label{EP} 
\end{eqnarray}
since the auxiliary work-performing system returns to its initial state (hence no change in its
entropy $\Delta S=0$) and the heat reservoirs are assumed to undergo a quasi-static heat exchange
while preserving their temperature. The fact that the total entropy cannot decrease, $\Delta S_{tot}
\ge 0$, is equivalent to the statement that efficiency is bounded by Carnot efficiency, $\eta \le
\eta_C$. The latter is reached for a reversible process, $\Delta S_{tot}=0$, which can only be achieved 
for a quasi-static transformation of the system implying infinitely slow Carnot cycles.
 
While the concept of Carnot efficiency is of paramount importance in the derivation of
thermodynamics, its practical implications are more limited: to reach the reversible limit 
one needs in principle to work with infinitely slow cycles.
Hence, the power of such a thermal machine is zero. This leaves open the question of efficiency at
finite power. Although this issue was first addressed by Chambadal \cite{Chambadal} and Novikov
\cite{Novikov}, it is often associated with the later work of Curzon and Ahlborn (CA) \cite{CA}.
Using an approximate analysis of a finite-time Carnot cycle, they observed that the power goes through 
a maximum, and that the corresponding efficiency at maximum power is given by the appealing expression 
\begin{eqnarray}
\eta_{CA}=1-\sqrt{\frac{T_c}{T_h}} \label{CAeff}.
\end{eqnarray}
Unfortunately, the CA efficiency turns out to be neither an
exact nor a universal result, and it is neither an upper nor a lower bound \cite{ftt}. Yet it
describes the efficiency of actual thermal plants very well \cite{CA,callen,plant}, and is reasonably close to the efficiency at maximum power for several model systems \cite{devos85,bejan1996,Hernandez,schmiedl08,then08,Okuda,Tu08,Allahverdyan,EspoEPL09a,EspoPRB09,EspoPRE10,Segal10}. 
How does this agreement come about?

As a first explanation, we note that  the underlying time-reversibility of the laws of physics under some
conditions implies  universal properties for the efficiency at maximum power. More precisely, let us
consider the expansion of  the efficiency at maximum power in terms of the Carnot efficiency
$\eta_{C}$. For CA efficiency one has $\eta_{CA}=1-\sqrt{1-\eta_{C}}=\eta_{C}/2+\eta_{C}^2/8+
\cdots$.  It was proven from the symmetry of the Onsager coefficients that the coefficient $1/2$ is
actually an upper bound for the linear response at maximum power, and that the bound is reached for
strong coupling between the heat- and the work-performing fluxes \cite{VdB05}. Using the equivalent
of Onsager symmetry at the level of nonlinear response, one can show  that the coefficient of
$\eta_{C}^2$ is also universal, i.e., equal to $1/8$, for strongly coupled fluxes possessing 
in addition a left-right symmetry \cite{EspoPRL09}. 
 
In this letter, we further clarify the special status of CA-efficiency: it turns out to be an exact property for Carnot machines operating under conditions of low, symmetric dissipation. The argument is very simple, as can be expected from its claim of generality.
Our starting point is a Carnot engine which operates under reversible conditions when the durations
of the cycles become very large, i.e., when the system always remains infinitesimally close to equilibrium all along the cycle. While in contact with the hot reservoir, the work-performing auxiliary system absorbs an amount of heat $Q_h$, resulting in a system entropy change $\Delta S = Q_h/T_h$. During the heat exchange with the cold reservoir, the entropy of the system returns to its original value, decreasing by an amount $-\Delta S = Q_c/T_c$. From the equality $Q_h/T_h=-Q_c/T_c$ we recover Carnot efficiency $\eta=1+Q_c/Q_h=1-T_c/T_h$.
We next consider finite-time cycles which move the engine away from the reversible regime.

Let $\tau_c$ ($\tau_h$) be the time durations during which the system is in contact with the cold (hot) reservoir along a cycle. In the weak dissipation regime, the system relaxation is assumed to be fast compared to $\tau_h$ and $\tau_c$. 
The entropy production per cycle along the cold (hot) part of the cycle is expected to behave as
$\Sigma_c/\tau_c$ ($\Sigma_h/\tau_h$) since the reversible regime is approached in the limits
$\tau_h \rightarrow \infty$ and $\tau_c \rightarrow \infty$ (for a further comment on this assumption
see \cite{Discussion1}). As a result, the amount of heat per cycle entering the system from the cold (hot) reservoir will be 
\begin{eqnarray}
Q_c=T_c \left (-\Delta S  -\frac{\Sigma_c}{\tau_c}+\dots \right) \nonumber\\ 
Q_h=T_h \left (\Delta S  -\frac{\Sigma_h}{\tau_h}+\dots \right). \label{ExpandEP}
\end{eqnarray}
Note that we did not specify the details of the procedure by which we deviate from the reversible
scenario. This information is contained in the coefficients $\Sigma_c$ and $\Sigma_h$. They express
how dissipation increases as one moves away from the reversible limit. We also do not need to assume 
that the temperature difference between $T_c$ and $T_h$ is small, hence the expansion is not limited 
to the linear response regime.

\begin{table*}
\center
\caption{Theoretical bounds and observed efficiency $\eta_\text{obs}$ of thermal plants\label{TABLE}}
\begin{tabular}{lcccccc}
\hline
Plant                               &$T_h(K)$&$T_c(K)$&$\eta_C$&$\eta_-$&$\eta_+$&$\eta_\text{obs}$\\
\hline \hline 
Doel 4 (Nuclear, Belgium)\cite{plant}      &$566$&$283$&$.5$  &$.25$&$.33$&$.35$\\ 
Almaraz II (Nuclear, Spain)\cite{plant}    &$600$&$290$&$.52$ &$.26$&$.35$&$.34$\\
Sizewell B (Nuclear, UK)\cite{plant}       &$581$&$288$&$.5$  &$.25$&$.34$&$.36$\\
Cofrentes (Nuclear, Spain)\cite{plant}     &$562$&$289$&$.49$ &$.24$&$.32$&$.34$\\
Heysham (Nuclear, UK)\cite{plant}          &$727$&$288$&$.60$ &$.30$&$.43$&$.40$\\
West Thurrock (Coal,UK)\cite{callen}       &$838$&$298$&$.64$ &$.32$&$.48$&$.36$\\
CANDU (Nuclear,Canada)\cite{callen}        &$573$&$298$&$.48$ &$.24$&$.32$&$.30$\\
Larderello (Geothermal,Italy)\cite{callen} &$523$&$353$&$.32$ &$.16$&$.19$&$.16$\\
Calder Hall (Nuclear,UK)\cite{plant}       &$583$&$298$&$.49$ &$.24$&$.32$&$.19$\\
(Steam/Mercury,US)\cite{plant}             &$783$&$298$&$.62$ &$.31$&$.45$&$.34$\\
(Steam,UK)\cite{plant}                     &$698$&$298$&$.57$ &$.29$&$.40$&$.28$\\
(Gas Turbine, Switzerland)\cite{plant}     &$963$&$298$&$.69$ &$.35$&$.53$&$.32$\\
(Gas Turbine, France)\cite{plant}          &$953$&$298$&$.69$ &$.34$&$.52$&$.34$\\
\hline
\end{tabular}
\end{table*}

We now consider the power generated during this Carnot cycle. Using (\ref{ExpandEP}), we get
\begin{eqnarray}
P &=& \frac{-W}{\tau_h  + \tau_c} \label{Power}\\
&=& \frac{Q_h+Q_c}{\tau_h  + \tau_c} \nonumber \\ 
&=&\frac{(T_h-T_c) \Delta S - T_h \Sigma_h/\tau_h - T_c \Sigma_c/\tau_c}{\tau_h+\tau_c}.\nonumber
\end{eqnarray}
The maximum power is found by setting the derivatives of $P$ with respect to $\tau_h$ and $\tau_c$ equal to zero. We find a unique physically acceptable solution at
\begin{eqnarray}
\tau_h = 2\frac{T_h \Sigma_h}{(T_h-T_c) \Delta S} \left(1+ \sqrt{\frac{T_c \Sigma_c}{T_h \Sigma_h}} \right) \nonumber\\
\tau_c = 2\frac{T_c \Sigma_c}{(T_h-T_c) \Delta S} \left(1+ \sqrt{\frac{T_h \Sigma_h}{T_c \Sigma_c}} \right).\label{OptTime}
\end{eqnarray}
Using (\ref{ExpandEP}) with (\ref{OptTime}) in the efficiency (\ref{EffiDef}) leads to the main
result of this paper, namely the following expression for the efficiency at maximum power:
\begin{eqnarray}
\eta^* = \frac{\eta_C \left(1+\displaystyle\sqrt{\frac{T_c \Sigma_c}{T_h \Sigma_h}} \right)}{ \left(1+\displaystyle\sqrt{\frac{T_c \Sigma_c}{T_h \Sigma_h}} \right)^2 
+ \displaystyle\frac{T_c}{T_h} \left(1-\frac{\Sigma_c}{\Sigma_h}\right) } \label{EffFinal}.
\end{eqnarray}
This result was previously obtained by Schmiedl and Seifert using a Fokker-Plank formulation of stochastic thermodynamics \cite{schmiedl08}. We present it here in a broader context by arguing that the expansion (\ref{ExpandEP}) is generic in the weak dissipation limit if a reversible long time limit exists.
For symmetric dissipation, $\Sigma_h=\Sigma_c$, we recover the Curzon-Ahlborn efficiency:
\begin{eqnarray}
\eta^*=\eta_{CA} = 1-\sqrt{\frac{T_c}{T_h}}=1-\sqrt{1-\eta_C}.
\end{eqnarray}
We also note that (\ref{EffFinal}) can be expanded in $\eta_C$ as
\begin{eqnarray}
\eta^* = \frac{\eta_C}{2} + \frac{\eta_C^2}{4+4\sqrt{\Sigma_c/\Sigma_h}} + {\cal O}(\eta_C^3)
\end{eqnarray}
The coefficient of the second order term lies between 0 and $1/4$ and
for symmetric dissipation we recover the $1/8$, as discussed in \cite{EspoPRL09}.
Symmetric dissipation for time-dependent cycles is thus similar to the left-right 
symmetry on the fluxes [see Eq. (20) of Ref. \cite{EspoPRL09}] which is required 
to recover the universal value of the quadratic coefficient for steady-state problems.

\begin{figure}[h]
\rotatebox{0}{\scalebox{0.48}{\includegraphics{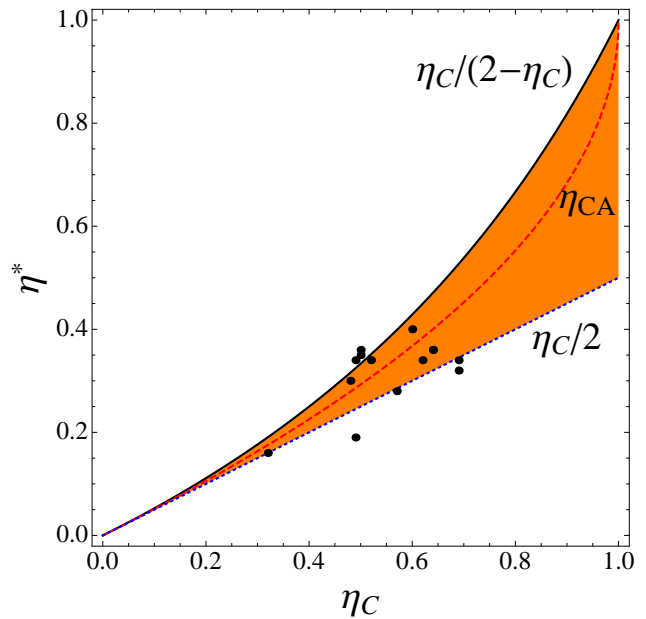}}} 
\caption{(Color online) Efficiency at maximum power as a function of $\eta_C$. The upper and lower bounds of the efficiency given by Eq. (\ref{BoundedEff}) are denoted by a black full line and a blue dotted line, respectively. The Curzon-Ahlborn efficiency is the red dashed line. The dots represent the observed efficiencies of the various thermal power plants reported in Table \ref{TABLE}. Observed efficiencies above and below the bounds could result from power plants not operating at maximum power.}  
\label{Fig1}
\end{figure}

We now turn to the main focus of the result (\ref{EffFinal}). In the limits $ \Sigma_c/ \Sigma_h \rightarrow 0$ and $\Sigma_c/ \Sigma_h\rightarrow \infty$, the efficiency at maximum power converges to the upper bound $\eta_{+}=\eta_C/(2-\eta_C)$ and to the lower bound $\eta_{-}=\eta_C/2$, respectively:
\begin{eqnarray}
\frac{\eta_C}{2} \equiv \eta_- \le \eta^* \le \eta_+ \equiv \frac{\eta_C}{2-\eta_C}. \label{BoundedEff}
\end{eqnarray}
In Fig. \ref{Fig1} we plot the efficiency (\ref{EffFinal}) as a function of $\eta_C$ for different
values of $\Sigma_c/\Sigma_h$, including the upper and lower bounds (\ref{BoundedEff}). 
We note that these bounds were previously derived by assuming a specific form of heat transfers in \cite{Yan89}.
The upper bound $\eta_+$, which is reached in the completely asymmetric limit $\Sigma_c/ \Sigma_h\rightarrow
0$, is particularly interesting. 
It coincides with a reported universal upper bound that was
derived in \cite{Gaveau} (cf. Eq. (16)) using a very different approach. It also agrees with the 
upper bound obtained by optimizing with respect to the temperature of the hot reservoir \cite{bejan1976}.
Finally, it also arises in a model for the Feynman ratchet \cite{Hernandez} (cf. Eq. (25)).

In order to identify the regime of operation of a particular engine other than via the ratio of coefficients 
$\Sigma_c/ \Sigma_h$, we evaluate the ratio of the contact times at maximum power:
\begin{eqnarray}
\frac{\tau_c}{\tau_h} 
=\sqrt{\frac{T_c\Sigma_c}{T_h\Sigma_h}}.
\end{eqnarray}
We conclude that symmetric dissipation corresponds to the case when this ratio is equal to the square root of the ratio of the temperatures,
\begin{eqnarray}
\displaystyle\frac{\tau_c}{\tau_h} = \displaystyle\sqrt{\frac{T_c}{T_h}},
\end{eqnarray}
whereas maximum and minimum efficiency are reached for the highly asymmetric cases
\begin{equation}
\frac{\tau_c}{\tau_h} \rightarrow 0 \quad \text{and} \quad \frac{\tau_c}{\tau_h} \rightarrow \infty.
\end{equation}

In conclusion, we have presented a simple and general argument for estimating efficiency of a
thermal engine at maximum power. The main bonuses of this analysis are the derivation of the Curzon
Ahlborn efficiency in the case of symmetric dissipation, and the prediction of an upper and lower
bound reached in the limits of extremely asymmetric dissipation. While actual plants usually operate
under steady state conditions rather than as a Carnot cycle, and  while the assumptions of low
dissipation and maximum power may not hold, one feels compelled to compare the upper and lower
bounds with observed efficiencies, as is done in Table \ref{TABLE} and in Fig. \ref{Fig1}. \\

\acknowledgments

M. E. is supported by the Belgian Federal Government (IAP project ``NOSY") and by the 
European Union Seventh Framework Programme (FP7/2007-2013) under grant agreement 256251.
This research is supported in part by the NSF under grant PHY-0855471.



\begin{thebibliography}{0}

\bibitem{callen}
H. B. Callen, {\it Thermodynamics and an Introduction to Thermostatistics} (Wiley, 2 ed., 1985).

\bibitem{Chambadal}
P. Chambadal, {\it Les Centrales Nucléaires}, (Armand Colin, Paris, 1957).

\bibitem{Novikov}
I. I. Novikov, Atomic Energy, {\bf 3}, 1269 (1957);
J. Nuclear Energy II, 7, 125 (1958).

\bibitem{CA}
F. Curzon and B. Ahlborn, Am. J. Phys. 43, 22 (1975).

\bibitem{plant}
A. Bejan, {\it Advanced Engineering Thermodynamics} (Wiley, New York, 1997); p. 377.

\bibitem{ftt}
K. H. Hoffmann, J. Burzler and S. Schubert, J. Non-Equilib. Thermodyn {\bf 22}, 311 (1997);
R. S. Berry, V. A. Kazakov, S. Sieniutycz, Z. Szwast, and A. M. Tsvilin, 
\textit{Thermodynamic Optimization of Finite-Time Processes} (Wiley, Chichester, 2000);
P. Salamon, J. D. Nulton, G. Siragusa, T. R. Andersen, and A. Limon, Energy {\bf 26}, 307 (2001).

\bibitem{devos85}
A. De Vos, Am. J. Phys. {\bf 53}, 570 (1985).

\bibitem{bejan1996}
A. Bejan, J. Appl. Phys. {\bf 79}, 1191 (1996).

\bibitem{Hernandez}                       
B. JimenezdeCisneros and A. C. Hernandez, Phys. Rev. Lett. {\bf 98}, 130602 (2007).  

\bibitem{schmiedl08}
T. Schmiedl and U. Seifert, Europhys. Lett. {\bf 81}, 20003 (2008).

\bibitem{then08}
H. Then and A. Engel, Phys. Rev. E {\bf 77}, 041105 (2008).

\bibitem{Okuda}
Y. Izumida and K. Okuda, EPL {\bf 83}, 60003 (2008);
Phys. Rev. E {\bf 80}, 021121 (2009);
Prog. Theor. Phys. Suppl. {\bf 178}, 163 (2009);
arXiv:1006.2589 .

\bibitem{Tu08}
Z. C. Tu, J. Phys. A {\bf 41}, 312003 (2008);

\bibitem{Allahverdyan}
A. E. Allahverdyan, R. S. Johal, and G. Mahler, Phys.
Rev. E 77, 041118 (2008).

\bibitem{EspoEPL09a}
M. Esposito, K. Lindenberg, and C. Van den Broeck, Europhys. Lett. {\bf 85}, 60010 (2009).

\bibitem{EspoPRB09}
B. Rutten, M. Esposito, B. Cleuren, Phys. Rev. B {\bf 80}, 235122 (2009).

\bibitem{EspoPRE10}
M. Esposito, R. Kawai, K. Lindenberg, and C. Van den Broeck, Phys. Rev. E {\bf 81}, 041106 (2010).

\bibitem{Segal10} 
Y. Zhou and D. Segal, Phys. Rev. E {\bf 82}, 011120 (2010).

\bibitem{VdB05}
C. Van den Broeck, Phys. Rev. Lett. 95, 190602 (2005); Adv. Chem. Phys. 135, 189 (2007).

\bibitem{EspoPRL09}
M. Esposito, K. Lindenberg, and C. Van den Broeck, Phys. Rev. Lett. {\bf 102}, 130602 (2009).

\bibitem{Discussion1}
In general, reconnecting the system to one of the reservoirs will induce 
dissipation because the system is not in equilibrium with this reservoir. 
This leads to an additional contribution to the entropy production in 
(\ref{ExpandEP}) which can have a complicated time dependence and does 
not vanish in the very slow cycle limit. This unavoidably results in a 
decrease of the efficiency at maximum power. By considering cycles 
with a reversible limit, we assume this dissipation to be negligible or absent. 
This assumption is reasonable in large systems but might require a large number 
of external control parameters for small systems (see for example \cite{sekimoto}). 
One can then use a calculation similar to the adiabatic theory in quantum mechanics 
to show that the entropy production related to the change of the system during the 
times $\tau$ while in contact with the reservoir behaves as $1/\tau$.

\bibitem{sekimoto}
K. Sato, K. Sekimoto, T. Hondou and F. Takagi, Phys. Rev. E {\bf 66}, 016119 (2002).
 
\bibitem{Gaveau}                       
B. Gaveau, M. Moreau and L. S. Schulman, Phys. Rev. Lett. {\bf 105}, 060601 (2010).

\bibitem{bejan1976}
A. Bejan and H. M. Paynter, {\it Solved Problems in Thermodynamics}  (MIT Press, Cambridge, MA, 1976), problem VII-D.

\bibitem{Yan89}
L. Chen and Z. Yan, J. Chem. Phys. {\bf 90}, 3740 (1989). 

\bibitem{Hernandez}
S. Velasco, J. M. M. Roco, A. Medina and A. Calvo Hern\'andez, J. Phys. D: Appl. Phys. {\bf 34}, 1000 (2001).

\end{thebibliography}
\end{document}